\documentclass[a4paper,fleqn]{cas-dc}

\usepackage[numbers]{natbib}
\usepackage{algorithm}
\usepackage{algorithmicx}
\usepackage{algpseudocode}
\usepackage{lipsum}
\usepackage{hyperref}
\usepackage{tikz}
\usepackage{amssymb}
\usepackage{caption}
\usepackage{tablefootnote}
\usepackage{longtable}
\usepackage{graphicx}
\usepackage{array}
\usepackage{hyperref}
\usepackage{booktabs}
\usepackage{placeins}
\usepackage{float}
\restylefloat{table}
\def\tsc#1{\csdef{#1}{\textsc{\lowercase{#1}}\xspace}}
\tsc{WGM}
\tsc{QE}
\tsc{EP}
\tsc{PMS}
\tsc{BEC}
\tsc{DE}
\begin{document}
\let\WriteBookmarks\relax
\def\floatpagepagefraction{1}
\def\textpagefraction{.001}

\shorttitle{BaCLNS: A toolbox for fast and efficient control of Linear and Nonlinear Control Affine Systems}

\shortauthors{Folorunsho et~al.}

\title [mode = title]{BaCLNS: A toolbox for fast and efficient control of Linear and Nonlinear Control Affine Systems}

\author[]{Samuel O. Folorunsho}
                        [orcid=0000-0001-6386-9190]
                        \cormark[1]

\ead{sof3@illinois.edu}

\cortext[cor1]{Corresponding author}


\credit{Methodology, Software, Data analysis and visualization, Data acquisition and curation, Manuscript preparation}

\affiliation[]{organization={University of Illinois, Urbana-Champaign},country={United States}}
\affiliation {organization={Center for Autonomous Construction and Manufacturing at Scale (CACMS)},country={United States}}

\author[]{William R. Norris} [orcid=0000-0002-4940-4458]%[style=chinese]
\credit{Manuscript review and editing, Project administration and supervision}

%\linenumbers

\begin{abstract}
Backstepping Control of Linear and Nonlinear Systems (BaCLNS) is a Python package developed to automate the design, simulation, and analysis of backstepping control laws for both linear and nonlinear control-affine systems. By providing a standardized framework, BaCLNS simplifies the process of deriving backstepping controllers, making this powerful control technique more accessible to engineers, researchers, and educators. The package handles complex system dynamics, ensuring robust stabilization even in the presence of significant nonlinearities. BaCLNS's modular design allows users to define custom control systems, simulate their behavior, and visualize the results, all within a user-friendly environment. The effectiveness of the package is demonstrated through a series of illustrative examples, ranging from simple linear systems to chaotic nonlinear systems, including the Vaidyanathan Jerk System, the pendulum and the Van der Pol Oscillator.
\end{abstract}

\begin{keywords}
Backstepping control \sep Nonlinear systems \sep Linear systems \sep Control-affine systems \sep Control law automation.
\end{keywords}

\maketitle

\section*{Metadata}
\label{}
Table~\ref{codeMetadata} and Table~\ref{executabelMetadata} show the code and software metadata respectively. 

\begin{table*}[!h]
\begin{tabular}{l p{7.8cm} p{7.8cm}}
%\Xhline{2\arrayrulewidth} % Top line
%\textbf{Code metadata description}\\
%\Xhline{2\arrayrulewidth} % Thicker line for the title row
Current code version & v0.1.1 \\
Permanent link to code/repository used for this code version & \url{https://github.com/sof-danny/BaCLNS} \\
Permanent link to Reproducible Capsule & None\\
Legal Code License & MIT License \\
Code versioning system used & Git \\
Software code languages, tools, and services used & Python \\
Compilation requirements, operating environments \& dependencies & Python v3.10 or higher, numpy, sympy, matplotlib \\
If available, link to developer documentation/manual & \url{https://github.com/sof-danny/BaCLNS} \\
Support email for questions & sof3@illinois.edu \\
%\Xhline{2\arrayrulewidth} % Bottom line
\end{tabular}
\caption{: Code metadata}
\label{codeMetadata} 
\end{table*}

\begin{table*}[!h]
\begin{tabular}{l p{7.8cm} p{7.8cm}}
%\Xhline{2\arrayrulewidth} % Top line
%textbf{(Executable) software metadata description} \\
%\Xhline{2\arrayrulewidth} % Thicker line for the title row
Current software version & v0.1.1 \\
Permanent link to executables of this version & \url{https://github.com/sof-danny/BaCLNS/releases/tag/v0.1.1} \\
Permanent link to Reproducible Capsule & None \\
Legal Software License & MIT License \\
Computing platforms/Operating Systems & Microsoft Windows, Linux, MacOS \\
Installation requirements \& dependencies & Python v3.10 or higher, numpy, sympy, matplotlib \\
If available, link to user manual -  
if formally \\published include a  reference 
to the publication in the reference \\ list & \url{https://github.com/sof-danny/BaCLNS/blob/main/README.md} \\
Support email for questions & sof3@illinois.edu \\
%\Xhline{2\arrayrulewidth} % Bottom line
\end{tabular}
\caption{: Software metadata}
\label{executabelMetadata} 
\end{table*}

\section{Motivation and Significance}

The complexity of modern dynamic systems, particularly those that exhibit nonlinear behavior, presents significant challenges for control engineers and researchers. Traditional control methods often struggle with nonlinearities, necessitating the development of more sophisticated strategies \cite{atherton1999limitations, choi2005performance}. One such approach is backstepping, a systematic and recursive control design technique that has proven effective for stabilizing nonlinear systems \cite{vaidyanathan2021introduction}.

Backstepping is particularly well-suited for \textit{control-affine systems} \cite{chitour2008singular}, which can be expressed in the general form:

\begin{equation}
\label{eqn1}
\dot{x} = f(x) + g(x)u,
\end{equation}

where \( x \in \mathbb{R}^n \) is the state vector, \( u \in \mathbb{R}^m \) is the control input, \( f: \mathbb{R}^n \rightarrow \mathbb{R}^n \) represents the system dynamics, and \( g: \mathbb{R}^n \rightarrow \mathbb{R}^{n \times m} \) represents the control influence on the system. The goal of backstepping is to design a control law \( u \) that stabilizes the system by recursively stabilizing each subsystem, ultimately leading to the stabilization of the entire system.

The backstepping approach begins by assuming that the system can be stabilized using a virtual control law at each step, then designing a lyapunov function that ensures asymptotic stability of the system's equilibrium point. Specifically, for a system described by the equation \ref{eqn1}, backstepping proceeds by first stabilizing a reduced-order system and then "stepping back" to incorporate additional state variables, until the full system is stabilized.

Despite its theoretical foundations, implementing backstepping control can be mathematically involved and its many steps often leads to errors which takes time to debug depending on the complexity of the system. This can be a barrier for researchers and engineers trying to use this control method. Despite this limitations, there is no known tool to automate this often laborious process.

Similar to other tools like do-mpc \cite{fiedler2023mpc}  and MPCTools \cite{aakesson2006mpctools} for analyzing and developing control laws for dynamic systems using Model Predictive Control, control.lqr \cite{pythoncontrol_lqr} for Linear Quadratic Regulators and simple-pid \cite{simplepid} for PID control, BaCLNS (Backstepping Control for Linear and Nonlinear Systems) was developed to address these challenges by providing a robust, flexible, and user-friendly framework for designing and simulating backstepping controllers. The package is specifically designed for control-affine systems of the form given in equation \ref{eqn1}, making it widely applicable across different domains where such systems are common, including robotics, aerospace, and mechanical systems. 

The package automates the derivation of control laws and provides simulation tools that reduce the time and effort required to develop and test control systems. The tool uses popular scientific computing libraries such as SymPy \cite{meurer2017sympy} and NumPy \cite{harris2020array} which ensures accessibility and compatibility with existing workflows, making advanced control techniques more approachable.

In recent years, there has been a significant surge in interest in applying neural networks to control applications \cite{zhang2024sparse, xie2024internal}. Typically, such applications require a substantial amount of training data to ensure that the model is robust. Generating this data involves creating numerous models and developing corresponding control laws, allowing the neural network to learn and predict effective control laws for previously unseen systems. Manually crafting these control law-system pairs is not only time-consuming but also prone to errors. In this context, BaCLNS can be invaluable by automating the creation of these control law-system pairs, thereby facilitating the training process.

BaCLNS has the potential to make a substantial impact on both academic research and practical engineering applications. For researchers, it offers a valuable tool for exploring new control strategies, validating theoretical models, and generating publishable results. Its versatility in handling both linear and nonlinear systems is particularly valuable for studying complex, real-world systems with high nonlinearities. In practical applications, BaCLNS enables engineers to rapidly prototype and test control systems, facilitating faster iteration and optimization of designs. Its availability as an open-source package encourages community engagement, enabling users to contribute to its development, share use cases, and advance the field of control systems engineering.

\section{Software Description}
BaCLNS (Backstepping Control for Linear and Nonlinear Systems) presents a simple solution for designing and simulating backstepping controllers across a variety of dynamic systems. The software architecture provides a modular and flexible structure, allowing users to easily integrate it into their existing workflows. Built in Python, BaCLNS leverages symbolic computation and numerical analysis libraries, enabling users to customize the control design process to suit both simple and complex system requirements. BaCLNS is equipped With robust functionality for automatic control law generation, simulation, and error analysis. The primary goal of BaCLNS is to provide a user-friendly and efficient framework for stabilizing linear and nonlinear control-affine systems. The package standardizes key functionalities such as system simulation, control law development, and performance visualization, ensuring that users can perform comprehensive evaluations and optimizations of their control systems faster. Illustrative examples and code snippets are included in this paper to demonstrate the application of BaCLNS to various control problems, further supported by detailed documentation available on the project's GitHub repository.

\subsection{System Architecture}

\begin{figure*}[h!]
    \centering
    \includegraphics[width=0.8\textwidth]{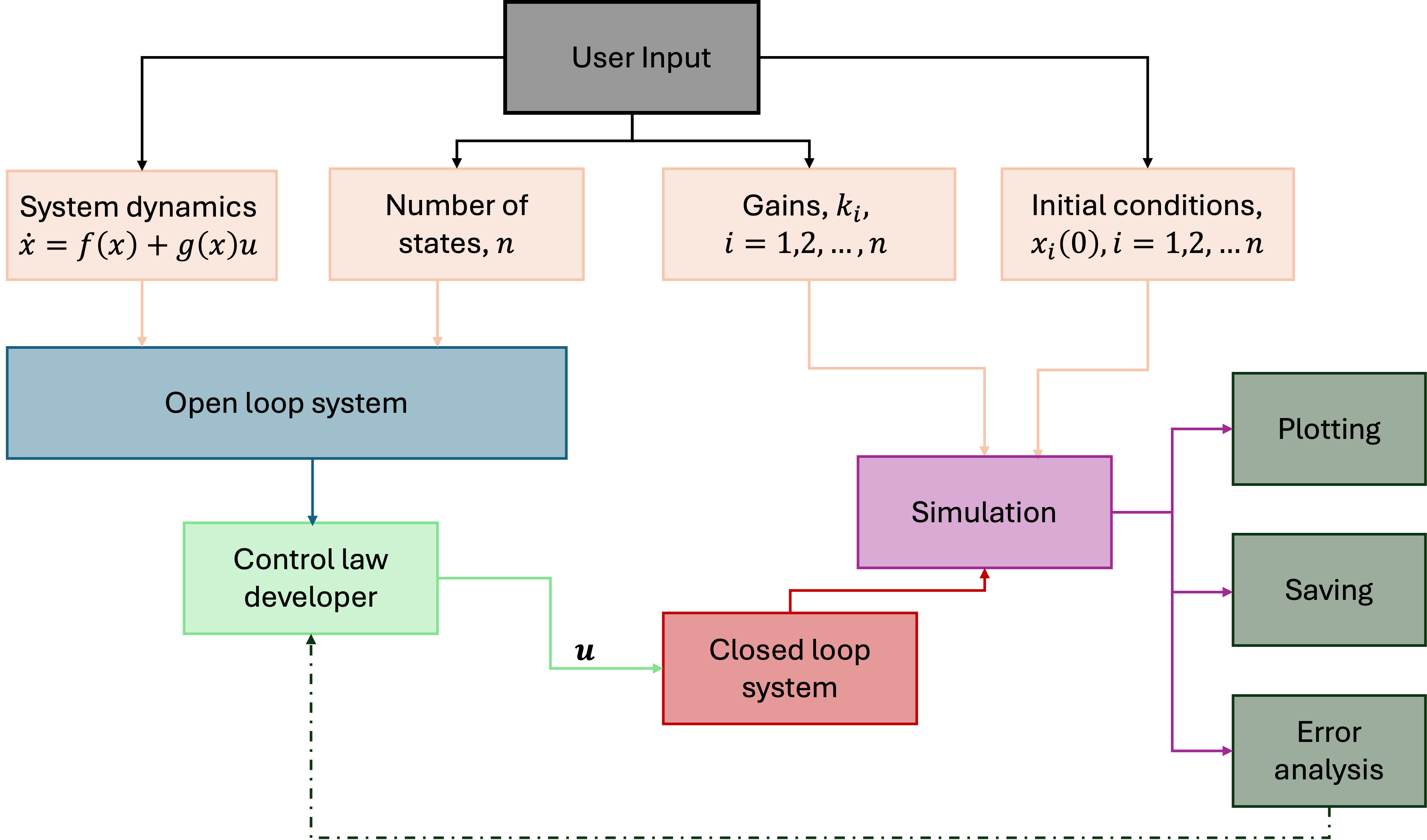}
    \caption{System architecture of BaCLNS.}
    \label{arch}
\end{figure*}

The overall workflow of BaCLNS is illustrated in Fig.~\ref{arch}, which outlines the key components and their interactions within the software.

The core components of the BaCLNS architecture include:

\begin{itemize}
    \item \textbf{User Input:} After installation using \textit{pip install BaCLNS}, the entry point of the workflow allow users to provide the system dynamics, the number of state variables, the control gains, and the initial conditions. These inputs form the foundation for the subsequent stages of control law development and simulation.
    
    \item \textbf{System Dynamics and State Definition:} The system dynamics are typically defined in the form of a control-affine system as in equation \ref{eqn1}. The number of states \( n \) and the specific form of the dynamics \( f(x) \) and \( g(x) \) are provided by the user.
    
    \item \textbf{Open-Loop System Response:} The open loop of the system is the state of the system with no feedback control. The output of the system dynamics and the number of states form the open loop system. Before designing a controller, users can simulate the open-loop response of the system, providing insights into the natural behavior of the system without any control input. 
    
    \item \textbf{Controller Developer:} The core of BaCLNS is the Controller Developer module, which generates the backstepping control law based on the provided system dynamics and state information following the steps outlines in Algorithm \ref{alg1}.
    The control law is derived recursively, stabilizing each subsystem sequentially, ultimately resulting in a control input \( u \) that stabilizes the entire system.

 \begin{algorithm}
\caption{Backstepping Control Design}\label{alg1}
\begin{algorithmic}[1]
\State \textbf{Input:} State-space dynamics, number of states $n$
\State \textbf{Output:} Control law $u$

\Function{BacksteppingControl}{$n$, $f_2, \dots, f_n$}
    \State Set $\phi_1(x_1) = -k_1 x_1$
    \State Define $V_1(x_1) = \frac{1}{2} x_1^2$ \cite{pukdeboon2011review}
    \State Compute $\dot{V}_1 = x_1 x_2$
    \State Define $z_1 = x_2 - \phi_1(x_1)$
    \State Define $\dot{z}_1 = \dot{x}_2 - \dot{\phi_1(x_1)}$
    \State Define $V_c(x_1, z_1) = V_1(x_1) + \frac{1}{2} z_1^2$
    \State Compute $\dot{V}_c = \dot{V}_1 + z_1 \dot{z}_1$
    \State Choose $u_i = -x_1 - k_2 z_1$ to ensure $\dot{V}_c < 0$
    \State Compute $u$ from $\dot{x}_2 = u_i - k_1$
    
    \If {$n > 2$}
        \State Repeat for $i=2,\dots,n-1$: Define $z_i$, $V_{c,i}$, and $\dot{V}_{c,i}$
        \State Compute $u_i$ iteratively to ensure $\dot{V}_{c,i} < 0$
    \EndIf
    
    \State \textbf{Return} $u$
\EndFunction
\end{algorithmic}
\end{algorithm}

    \item \textbf{Closed-Loop System Response and Simulation:} After the control law is generated, BaCLNS simulates the closed-loop response of the system. This simulation integrates the control law with the system dynamics using the gains and initial conditions provided by the user. The simulation allows users to observe how the system responds under control. It handles the numerical integration of the system's differential equations and provides the necessary data for further analysis.
    
    \item \textbf{Plotting, Saving, and Error Analysis:} BaCLNS includes tools for visualizing the simulation results, including state trajectories, control inputs, and errors over time. Users can generate plots to examine the system's behavior visually and save these plots for documentation or reporting purposes. Additionally, BaCLNS provides error analysis tools to quantify the performance of the controller by comparing the actual system states with the desired states.
\end{itemize}

\subsection{System Functionalities}

BaCLNS offers a range of functionalities designed to simplify the process of backstepping control design and analysis. The key functionalities include:

\begin{itemize}
    \item \textbf{Automatic Control Law Generation:} Users can input the system's state equations and number of states, and BaCLNS automatically derives the corresponding backstepping control law following Algorithm \ref{alg1}. This feature eliminates the need for manual derivation, which can be error-prone and time-consuming, particularly for complex systems.
    
    \item \textbf{Simulation of Dynamic Systems:} BaCLNS supports both open-loop and closed-loop simulations, enabling users to observe the behavior of their system under various scenarios. The simulation engine is customizable, allowing users to define their own time steps, initial conditions, and parameter values.
    
    \item \textbf{Plotting and Visualization:} The package includes built-in plotting functions that allow users to visualize state trajectories, control inputs, and errors over time. These plots can be saved automatically to a specified directory, making it easier to document and share results.
    
    \item \textbf{Error Analysis:} BaCLNS provides tools for error analysis, which are essential for assessing the performance of the control law. By comparing the actual system states with the desired states, users can quantify the effectiveness of their controller and identify areas for improvement.
    
    \item \textbf{Saving and Exporting Results:} Simulation results, including state trajectories, control inputs, and error data, can be saved in a variety of formats (e.g., JSON, CSV) for further analysis or inclusion in reports and publications.
\end{itemize}

\subsection{Sample Code Snippets}

BaCLNS is designed to be intuitive and easy to use. Below is an example of how to use BaCLNS to generate a backstepping control law for a simple linear system:

\begin{figure}[h]
    \centering
\includegraphics[width=0.45\textwidth]{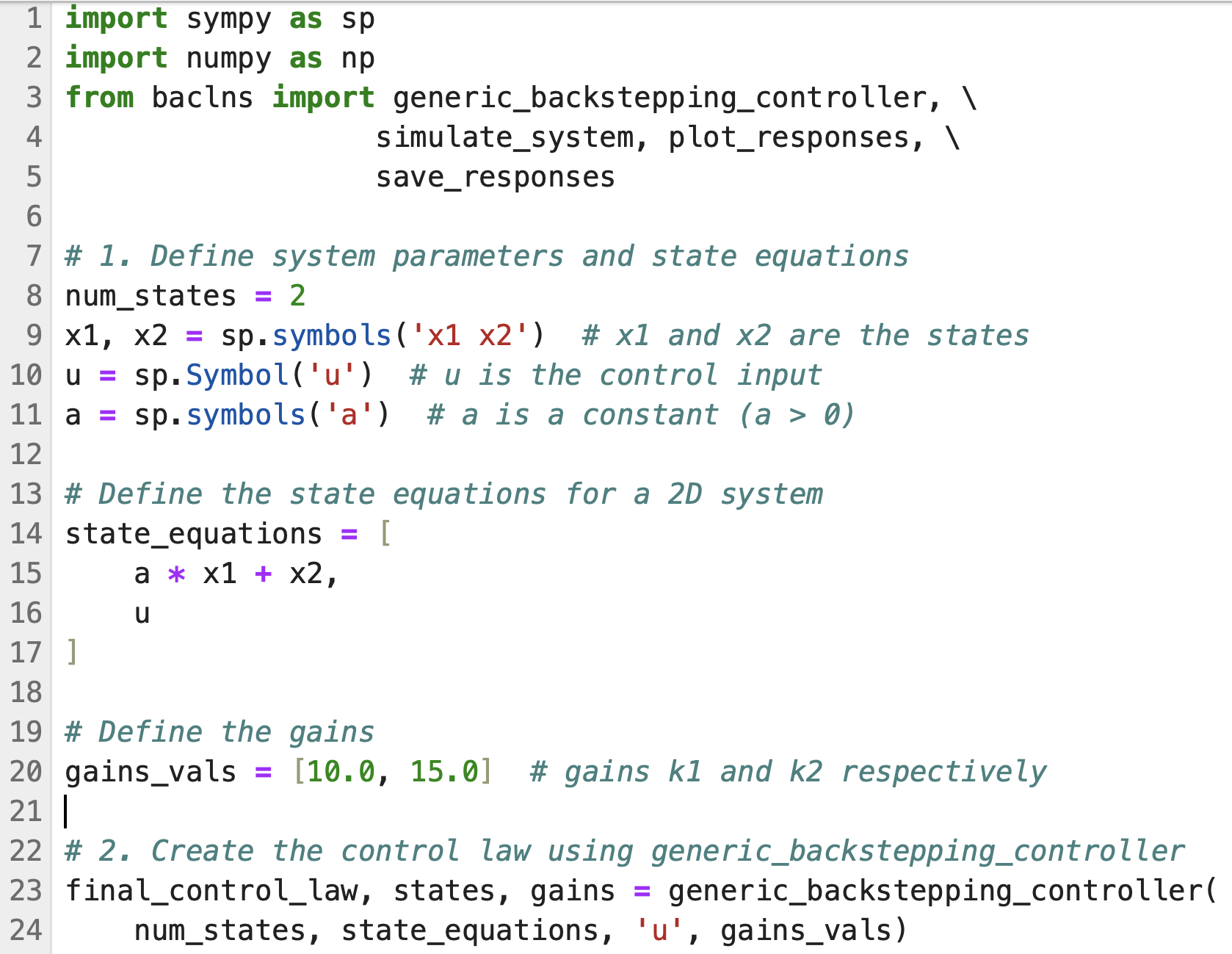}
\end{figure}

\begin{figure}[h]
    \centering
\includegraphics[width=0.45\textwidth]{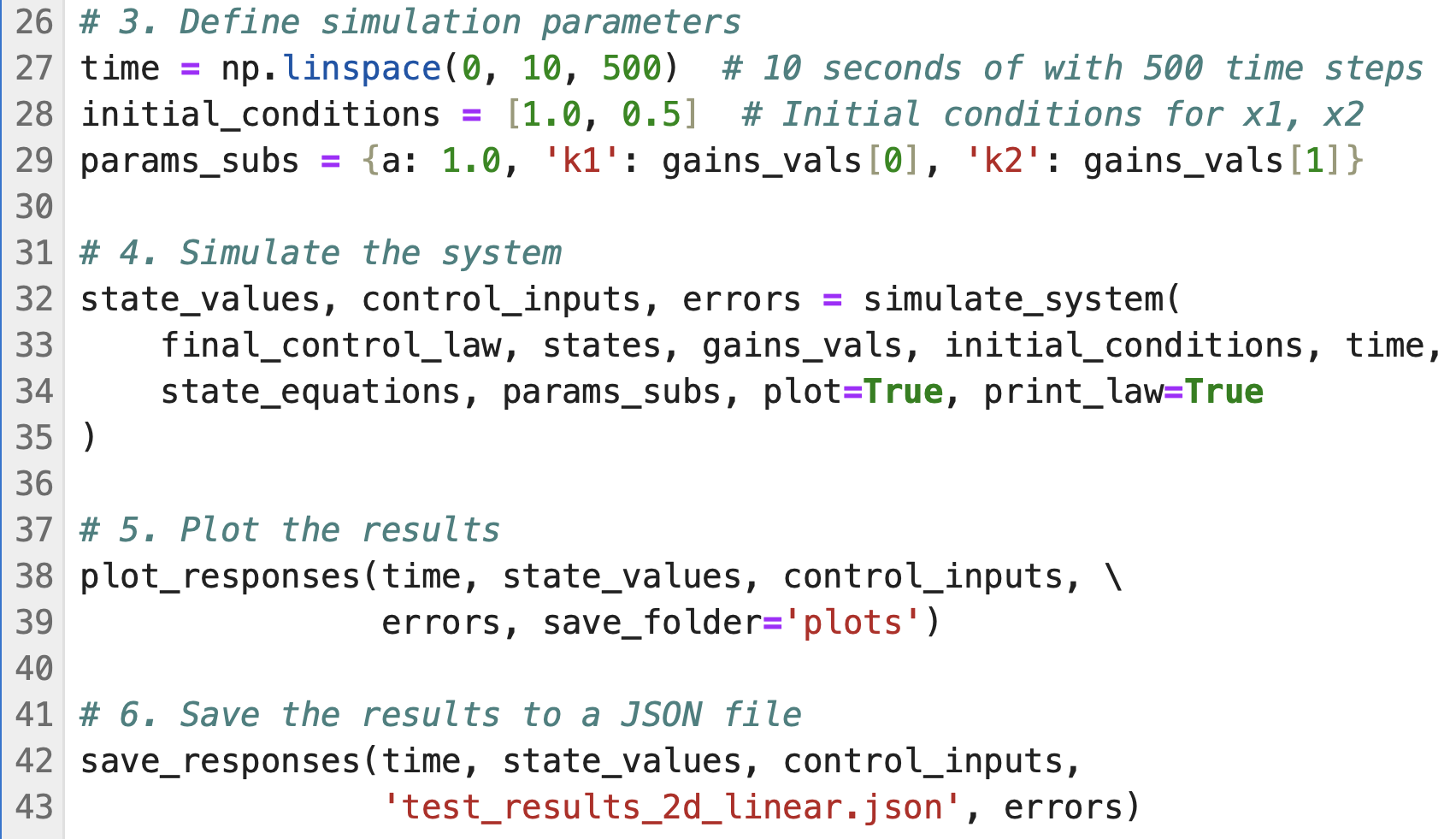}
\end{figure}

The flexibility and ease of use make it an ideal tool for both researchers and practitioners in the field of control systems engineering.

\section{Illustrative Examples}

In this section, the capabilities of BaCLiNS are demonstrated through a series of illustrative examples involving both linear and nonlinear systems. Each example includes the system dynamics, the control law generated by BaCLNS, the resulting state evolution demonstrating the stabilization of the system to the desired state (typically the origin) and the control input over time.

\subsection{2D Linear System}

The analysis begins with a simple 2D linear system, which serves as a basic example to illustrate the core functionality of BaCLiNS.

\textbf{State Equations:}
\[
\dot{x}_1 = a \cdot x_1 + x_2, \quad \dot{x}_2 = u
\]

\textbf{Control Law:}
BaCLNS generated the following control law to stabilize the system:
\[
u = -a \cdot k_1 \cdot x_1 - k_1 \cdot k_2 \cdot x_1 - k_1 \cdot x_2 - k_2 \cdot x_2
\]

\begin{figure}[h!]
    \centering
    \includegraphics[width=0.45\textwidth]{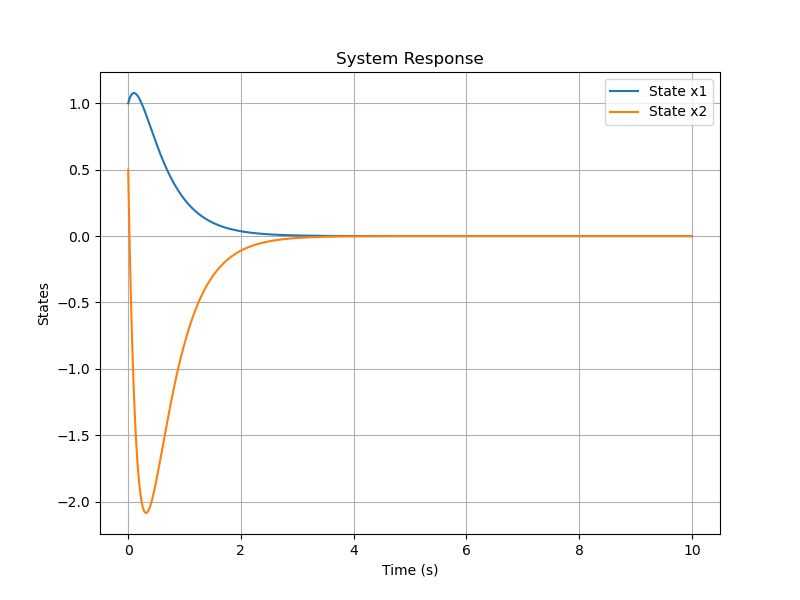}
    \caption{State evolution of the 2D linear system under the control law generated by BaCLNS.}
    \label{fig:2d_linear_state_evolution}
\end{figure}

\begin{figure}[h!]
    \centering
    \includegraphics[width=0.45\textwidth]{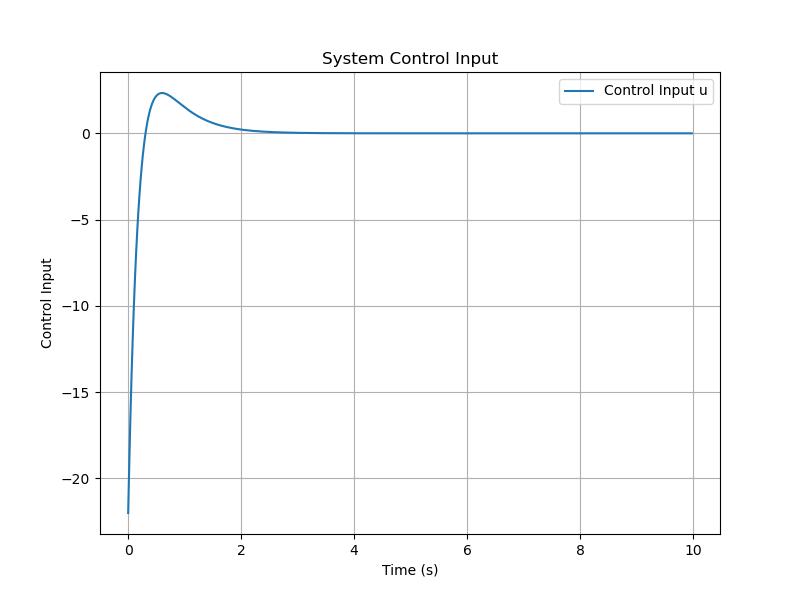}
    \caption{Control input for the 2D linear system over time}
    \label{2dlinearcontrol}
\end{figure}

\textbf{Simulation Results:}
Fig.~\ref{fig:2d_linear_state_evolution} shows the state evolution of the system under the influence of the generated control law Fig.~\ref{2dlinearcontrol}. The system states successfully converge to zero, demonstrating the effectiveness of the control law.

\subsection{3D Linear System}

Next, a 3D linear system is considered, extending the complexity of the control problem.

\textbf{State Equations:}
\[
\dot{x}_1 = a \cdot x_1 + x_2, \quad \dot{x}_2 = b \cdot x_3, \quad \dot{x}_3 = u
\]

\textbf{Control Law:}
The control law generated by BaCLNS for this system is:
\begin{align*}
u &= -a \cdot k_1 \cdot k_2 \cdot x_1 - b \cdot k_2 \cdot x_3 
    - k_1 \cdot k_2 \cdot k_3 \cdot x_1 \nonumber \\
  &\quad - k_1 \cdot k_2 \cdot x_2 - k_2 \cdot k_3 \cdot x_2 
    - k_3 \cdot x_3
\end{align*}

\begin{figure}[h!]
    \centering
    \includegraphics[width=0.45\textwidth]{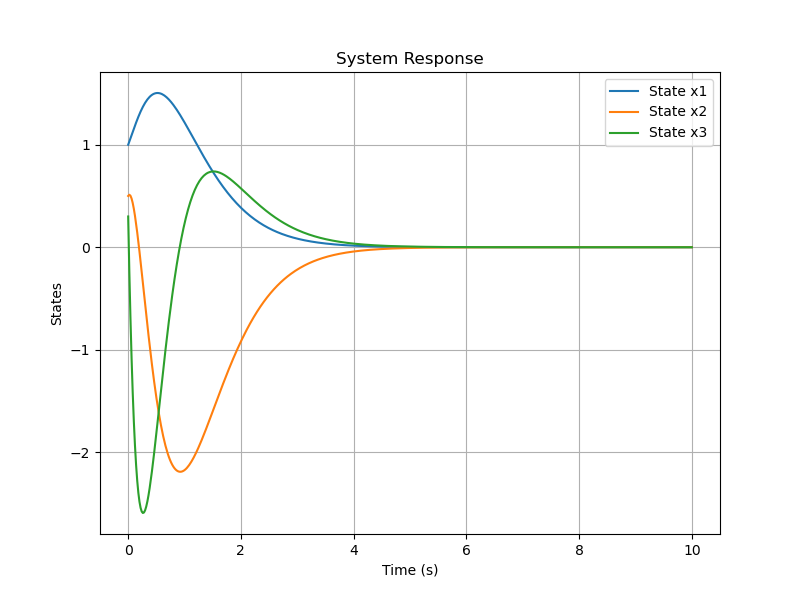}
    \caption{State evolution of the 3D linear system under the control law generated by BaCLNS.}
    \label{fig:3d_linear_state_evolution}
\end{figure}

\begin{figure}[h!]
    \centering
    \includegraphics[width=0.45\textwidth]{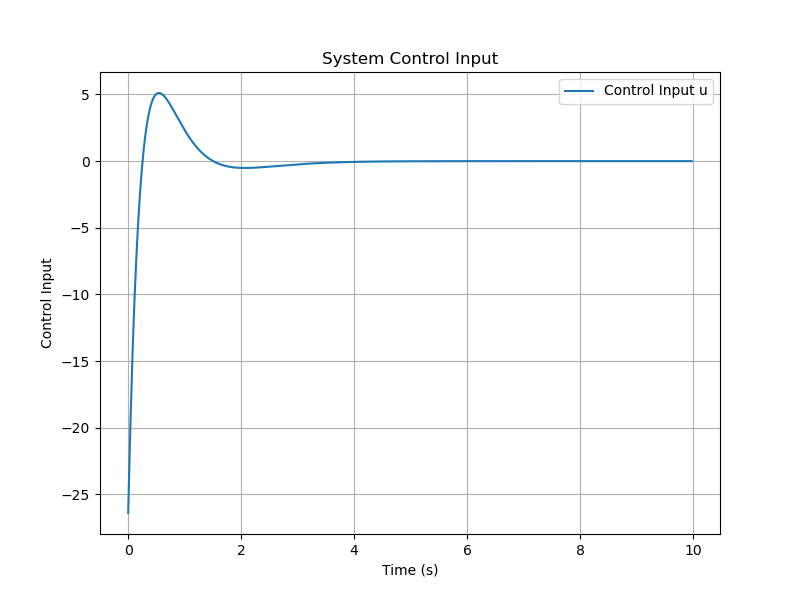}
    \caption{Control input for the 3D linear system over time}
    \label{3dlinearcontrol}
\end{figure}

\textbf{Simulation Results:}
The state evolution under this control law is shown in Fig.~\ref{fig:3d_linear_state_evolution}. The states converge to the origin based on inputs from the controller shown in Fig.~\ref{3dlinearcontrol}, indicating stabilization.

\subsection{2D Nonlinear System}

A 2D nonlinear system is now explored, introducing nonlinearity in the state equations and making the control problem more challenging.

\textbf{State Equations:}
\[
\dot{x}_1 = a \cdot x_1^2 + x_1^3 + x_2, \quad \dot{x}_2 = u
\]

\textbf{Control Law:}
BaCLNS generated the following control law:
\[
u = -a \cdot k_1 \cdot x_1^2 - k_1 \cdot k_2 \cdot x_1 - k_1 \cdot x_1^3 - k_1 \cdot x_2 - k_2 \cdot x_2
\]

\begin{figure}[h!]
    \centering
    \includegraphics[width=0.45\textwidth]{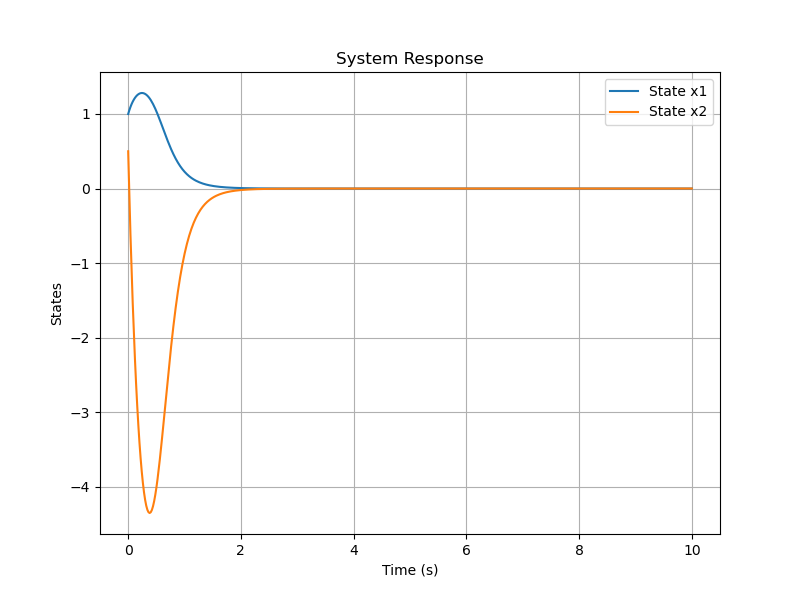}
    \caption{State evolution of the 2D nonlinear system under the control law generated by BaCLNS.}
    \label{fig:2d_nonlinear_state_evolution}
\end{figure}

\begin{figure}[h!]
    \centering
    \includegraphics[width=0.45\textwidth]{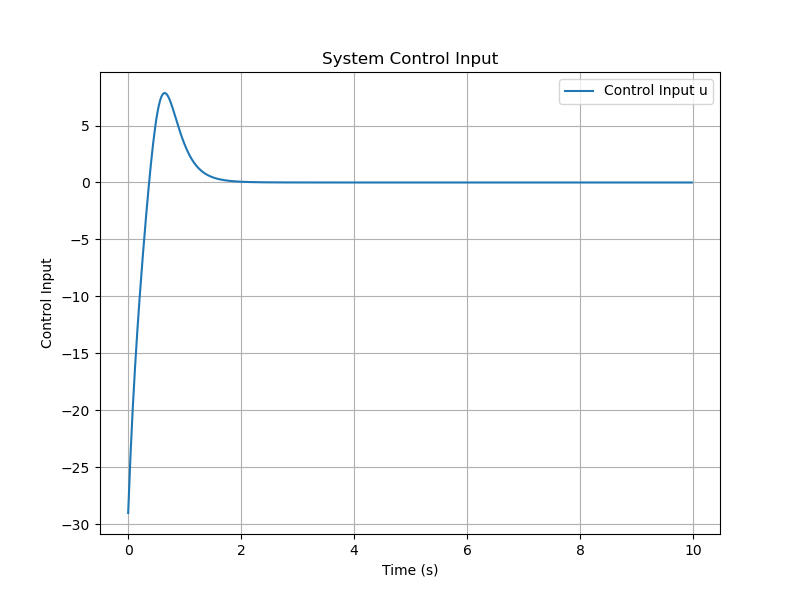}
    \caption{Control input for the 2D nonlinear system over time}
    \label{2dnonlinearcontrol}
\end{figure}

\textbf{Simulation Results:}
Fig.~\ref{fig:2d_nonlinear_state_evolution} shows the state trajectories for this nonlinear system and Fig.~\ref{2dnonlinearcontrol} shows the control input, demonstrating the system's stabilization.

\subsection{3D Nonlinear System (Vaidyanathan Jerk System)}

The Vaidyanathan Jerk System \cite{vaidyanathan2015analysis} is a well-known example of a chaotic system. Here, BaCLNS is used to stabilize it.

\textbf{State Equations:}
\[
\dot{x}_1 = x_2, \quad \dot{x}_2 = x_3, \quad \dot{x}_3 = a \cdot x_1 - b \cdot x_2 - c \cdot x_3 - x_1^2 - x_2^2 + u
\]

\textbf{Control Law:}
The control law derived by BaCLNS is:

\begin{align*}
u &= -a \cdot x_1 + b \cdot x_2 + c \cdot x_3 
- k_1 \cdot k_2 \cdot k_3 \cdot x_1 \nonumber \\
&\quad - k_1 \cdot k_2 \cdot x_2 
- k_2 \cdot k_3 \cdot x_2 \nonumber \\
&\quad - k_2 \cdot x_3 - k_3 \cdot x_3 
+ x_1^2 + x_2^2
\end{align*}

\begin{figure}[h!]
    \centering
    \includegraphics[width=0.45\textwidth]{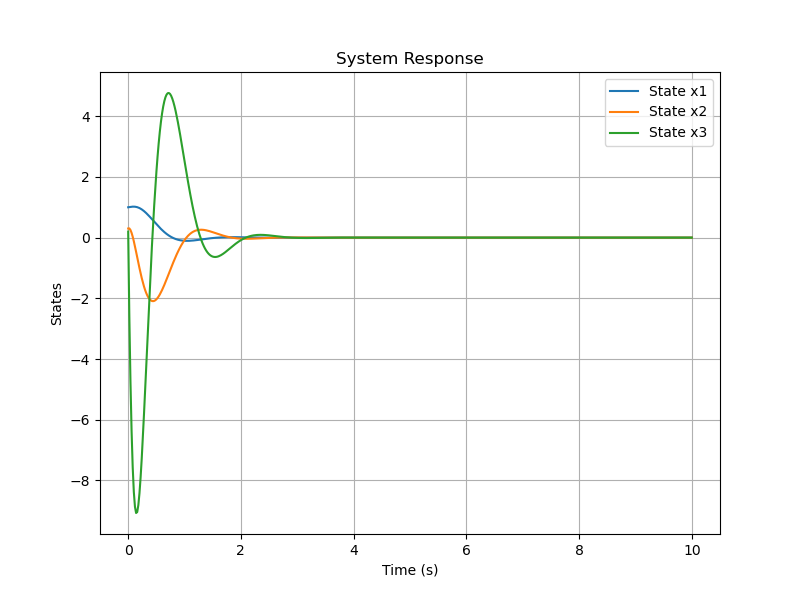}
    \caption{State evolution of the Vaidyanathan Jerk System under the control law generated by BaCLNS.}
    \label{fig:vaidyanathan_jerk_state_evolution}
\end{figure}

\begin{figure}[h!]
    \centering
    \includegraphics[width=0.45\textwidth]{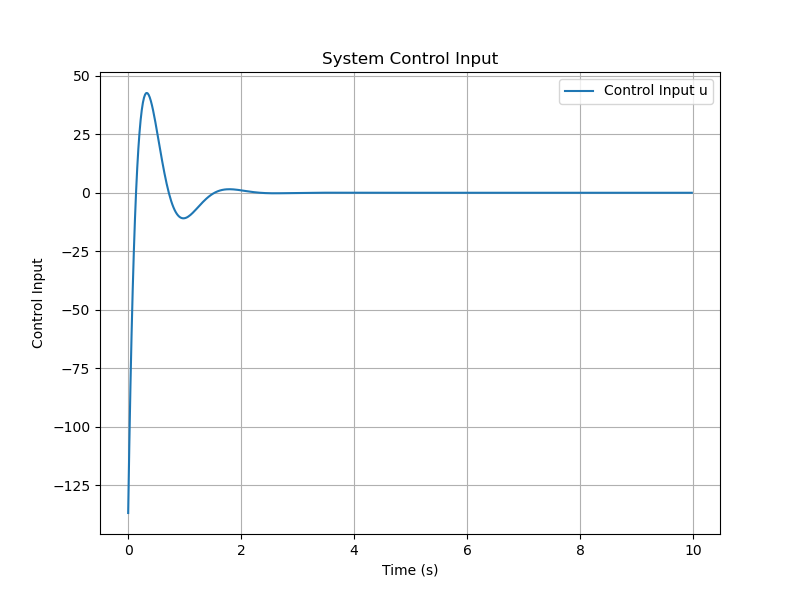}
    \caption{Control input for the Vaidyanathan Jerk System}
    \label{3dnonlinearcontrol}
\end{figure}

\textbf{Simulation Results:}
Fig.~\ref{fig:vaidyanathan_jerk_state_evolution} illustrates the state evolution, showing the effectiveness of the control law in stabilizing the system. The control input is shown in Fig.~\ref{3dnonlinearcontrol}.

\subsection{Simple Pendulum and the Van der Pol Oscillator}

The final examples are the classic simple pendulum \cite{nelson1986pendulum} and the Vander Pol Oscilator \cite{kanamaru2007van} systems. These are benchmark problems in control theory. It was necessary to test BaCLNS on classical real-world systems to demonstrate its practical applicability and effectiveness in stabilizing systems with well-known dynamics. The dynamics of the simple pendulum can be described by equation \ref{pend}:

\begin{equation}
\label{pend}
\begin{aligned}
\dot{x}_1 &= x_2, \\
\dot{x}_2 &= \frac{1}{ml^2} \left( u - b x_2 - mgl \sin(x_1) \right),
\end{aligned}
\end{equation}

where \(x_1 = \theta \) is the angle of the pendulum from the vertical, \(x_2 = \dot{\theta}\) is the angular velocity of the pendulum, $u$ is the control input (torque applied at the pivot), $m$: the mass of the pendulum, $l$: the length of the pendulum, $b$: the damping coefficient,$g$: the acceleration due to gravity.

Using the BaCLNS backstepping controller design algorithm, the following control law is produced:

\begin{equation*}
u = b x_2 + g l m \sin(x_1) - k_1 k_2 l^2 m x_1 - k_1 l^2 m x_2 - k_2 l^2 m x_2
\end{equation*}

\begin{figure}[h!]
    \centering
    \includegraphics[width=0.45\textwidth]{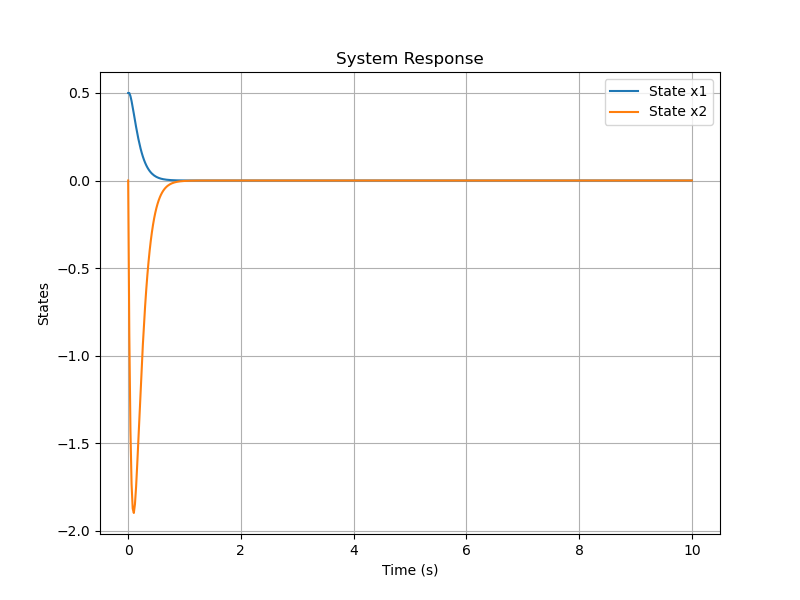}
    \caption{State evolution of the pendulum under the control law generated by BaCLNS.}
    \label{fig:pendulum_state_evolution}
\end{figure}

\begin{figure}[h!]
    \centering
    \includegraphics[width=0.45\textwidth]{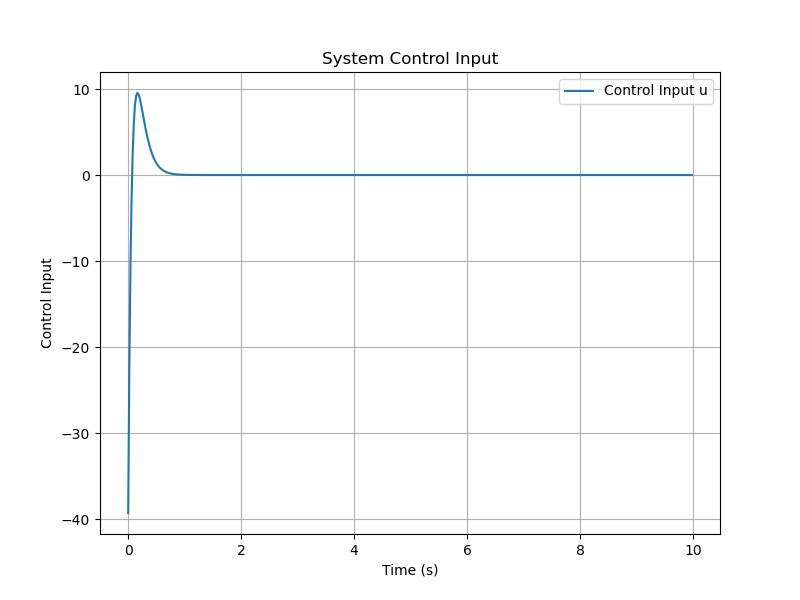}
    \caption{Control input for the Pendulum}
    \label{PendulumInput}
\end{figure}

\textbf{Simulation Results:}
Fig.~\ref{fig:pendulum_state_evolution} shows the state evolution, confirming that the control law stabilizes the pendulum. The control input is shown in Fig.~\ref{PendulumInput}.

The dynamics of the Van der Pol oscillator can be described by equation \ref{vdp}:

\begin{equation}
\label{vdp}
\begin{aligned}
\dot{x}_1 &= x_2, \\
\dot{x}_2 &= \mu(1 - x_1^2)x_2 - x_1 + u,
\end{aligned}
\end{equation}

where \(x_1\) is the position of the oscillator, \(x_2\) is the velocity, \(u\) is the control input, and \(\mu\) is a scalar parameter that controls the nonlinearity and the damping in the system.

Using the BaCLNS backstepping controller design algorithm, the following control law is produced:

\begin{equation*}
u = -k_1 k_2 x_1 - k_1 x_2 - k_2 x_2 + \mu x_1^2 x_2 - \mu x_2 + x_1
\end{equation*}

\begin{figure}[h!]
    \centering
    \includegraphics[width=0.45\textwidth]{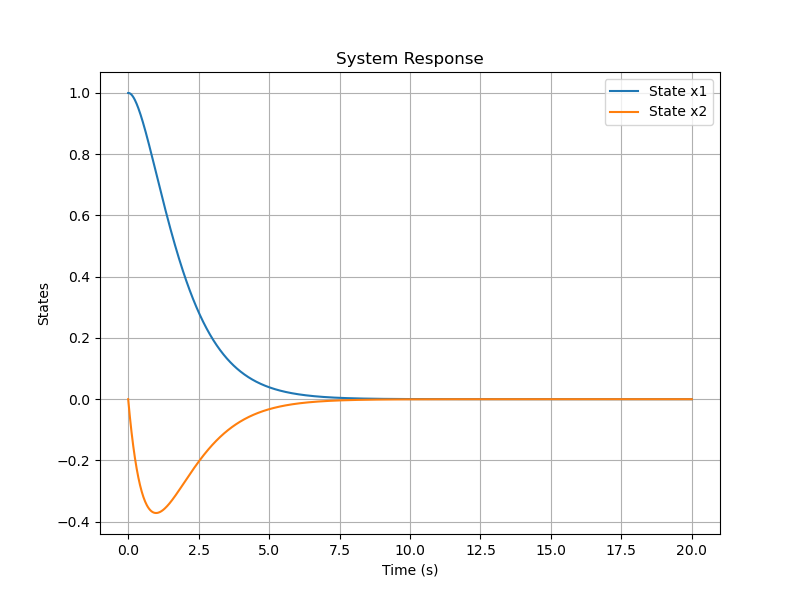}
    \caption{State evolution of the Van der Pol oscillator with \(\mu =1 \) under the control law generated by BaCLNS with.}
    \label{fig:vdp_state_evolution}
\end{figure}

\begin{figure}[h!]
    \centering
    \includegraphics[width=0.45\textwidth]{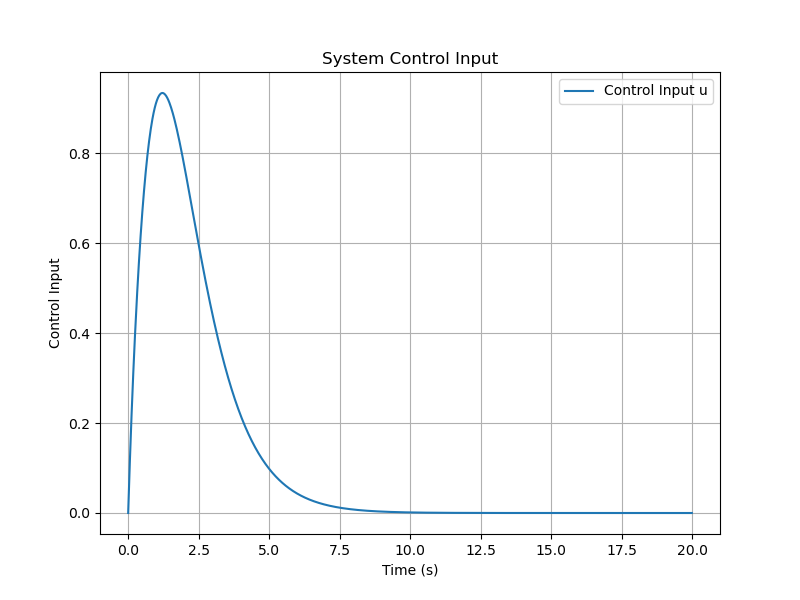}
    \caption{Control input for the Van der Pol oscillator}
    \label{vdpInput}
\end{figure}

\textbf{Simulation Results:}  
Fig.~\ref{fig:vdp_state_evolution} shows the state evolution, confirming that the control law stabilizes the Van der Pol oscillator. The control input is shown in Fig.~\ref{vdpInput}.

\section{Impact}
Control systems are integral to many areas of engineering and technology, providing the necessary mechanisms to manage the behavior of dynamic systems. The backstepping control technique has proven to be a robust method for stabilizing nonlinear systems, making it an important tool in the control systems engineer's toolkit \cite{marquez2003nonlinear}. Despite its importance, the process of deriving backstepping control laws, particularly for complex systems, can be tedious and prone to errors when done manually. BaCLNS addresses this challenge by automating the design, simulation, and analysis of backstepping control laws, thereby streamlining the entire control design process.

The introduction of BaCLNS significantly impacts the field of control systems engineering by offering a standardized approach to backstepping control design similar to other such tools in the domain. Traditionally, researchers and practitioners have relied on custom-built solutions tailored to specific systems, leading to inconsistencies and difficulties in reproducing results across different studies \cite{desrocherschallenges}. BaCLNS bridges this gap by providing a simple and consistent framework that can be applied to a wide range of linear and nonlinear systems. This standardization enhances the reliability of control system evaluations and ensures that results can be meaningfully compared across different studies and applications.

One of the major strengths of the BaCLNS package is its ability to handle both linear and nonlinear control-affine systems, making it applicable to a broad scope of control problems. Due to its automated approach,BaCLNS reduces the complexity involved in stabilizing complex systems, allowing engineers to focus on higher-level design and analysis tasks. The package’s versatility is further demonstrated through its application to various illustrative examples, ranging from simple linear systems to chaotic nonlinear systems like the Vaidyanathan Jerk System. These examples shows BaCLNS’s capability to effectively design control laws that stabilizes complex systems, thereby providing a powerful tool for both academic research and practical engineering applications.

BaCLNS promotes the adoption of backstepping control techniques by making them more accessible to a wider audience \cite{goodwin2001control}. The package is not only a valuable resource for control systems engineers but also for educators and students who are learning about advanced control techniques. Students have often times focused on using linearization approach for stability analyzes and control \cite{hou2016overview}. While this is useful for simple , predictable systems, their limitations become apparent when dealing with more chaotic systems with high nonlinearities. By providing a user-friendly interface and detailed documentation, BaCLNS facilitates the learning process and encourages the exploration of backstepping control in various educational and research settings.

Also, the output from BaLNS allows it to be easily integrated with existing control system analysis tools, allowing users to incorporate the package into their existing workflows without disruption. For example, the control law generated can be used in tools like MATLAB \cite{lee2023programming}, Robot Operating System (ROS) \cite{macenski2022robot} and Python-based simulation environments, to enhance their control system designs. The ability to export simulation results and control laws in various formats further extends the package’s utility, thereby enabling easy sharing and collaboration across different platforms and teams.

\section{Conclusion and Future Works}

In this paper, BaCLiNS, a Python package designed to automate the derivation, simulation, and analysis of backstepping control laws for both linear and nonlinear control affine systems, is introduced. BaCLNS provides a standardized and accessible framework for control systems engineers, researchers, and educators, significantly simplifying the process of designing and implementing backstepping controllers. Through a series of illustrative examples, the package's versatility and effectiveness in stabilizing a variety of dynamic systems were demonstrated, ranging from simple linear models to more complex and chaotic systems such as the Vaidyanathan Jerk System, the classic pendulum, and the Van der Pol Oscillator systems. The tool's ability to handle a wide range of systems, along with its user-friendly interface and detailed documentation, positions BaCLiNS as a valuable asset in the field of control systems engineering.

\par While BaCLiNS provides a robust foundation for backstepping control design, several enhancements and features are planned for future versions. One such feature is the integration of adaptive control techniques \cite{zhou2008adaptive}, which will allow the package to handle systems with uncertain or time-varying parameters more effectively. Additionally, support for higher-order sliding mode controllers \cite{shtessel2014sliding} is planned, which will expand the package's applicability to systems with significant disturbances or uncertainties. Another planned enhancement is the inclusion of more advanced visualization tools, such as 3D state-space trajectories and real-time control performance metrics, to provide users with deeper insights into system behavior during simulation.

\printcredits

\vspace{6pt} 
\begin{flushleft}
   \textbf{Acknowledgments}

    \vspace{4pt} 
    None

\end{flushleft}

%\bibliographystyle{ieee}
\bibliographystyle{IEEEtran}

%\bibliographystyle{cas-model2-names}

\bibliography{cas-refs}

\end{document}